# VARIABILITY OF ACTIVE GALACTIC NUCLEI FROM THE OPTICAL TO X-RAY REGIONS

C. MARTIN GASKELL[1,2] and ELIZABETH S. KLIMEK[3]

*Department of Physics & Astronomy, University of Nebraska, Lincoln, NE 68588-0111, USA*

Some of the progress in understanding the variability of active galactic nuclei (AGNs) from the optical to X-ray regions is reviewed. Although where there is a clear correlation between variations in the two regions, the optical lags the X-rays, simple reprocessing of the X-ray radiation to produce significant amounts of longer-wavelength continua seems to be ruled out. In a couple of objects where there has been correlated X-ray and optical variability, the amplitude of the optical variability has exceeded the amplitude of the X-ray variability. We suggest that the factor linking the X-ray and optical regions might not be irradiation, but accelerated particles striking matter (as in activity in the solar chromosphere). The diversity in optical/X-ray relationships at different times in the same object, and between different objects, could be explained by evolving differences in local geometry, and by changing directions of motion relative to our line of sight. Linear shot-noise models of the variability are ruled out; instead there must be large-scale organization of variability. Variability occurs on light-crossing timescales rather than viscous timescales and this probably rules out the standard Shakura-Sunyaev accretion disk. Instead, we believe that the main energy-generation mechanism is probably electromagnetic. The overall average continuum shape appears to be the same in both radio-loud and radio-quiet AGNs, strongly suggesting a similar origin to the continua. Radio-loud and radio-quiet AGNs have quite similar optical variability properties, and this suggests a common variability mechanism. Beaming effects could be significant in all types of AGN. Despite their extreme X-ray variability properties, our observations show that narrow-line Seyfert 1s (NLS1s) do not show extreme optical variability, and that their optical variability properties could well be similar to those of non-NLS1s.

## 1  OVERVIEW

The topic of variability of active galactic nuclei (AGNs) is a very large one, and the space available here does not permit a thorough review of this topic, which in recent years has generated hundreds of papers and much international

---

[1] Corresponding author. E-mail: *gaskell@astro.as.utexas.edu*
[2] Present address: Astronomy Department, University of Texas, Austin, TX 78712-0259, USA.
[3] Present address: Astronomy Department, New Mexico State University, Las Cruces, NM 88003-0001, USA

collaboration. We therefore just give a brief, and necessarily selective, review of a little of the history, a listing of what we consider to be some of the important questions to be answered, a report on some recent work we have been involved in, comments on a few results we think worth noting, our current feelings on what we think variability has been telling us, and some suggestions for future collaborations. For a longer general review of AGN variability see Ulrich, Maraschi, & Urrey (1997). For an earlier review of X-ray variability alone up to circa 1992 see Mushotzky, Done, Pounds (1993).

## 2  EARLY HISTORY

Probably all AGNs vary at all wavelengths, and this variability has been recognized for a long time. In fact, optical variability was discovered before the true nature of AGNs was appreciated. In 1956 A. Deutsch at the Pulkova Observatory reported that the magnitude of the nucleus of NGC 5548 appeared to vary by about a magnitude. At the time it was believed that the continuous optical emission from the nuclei of galaxies was entirely due to the emission from many millions to billions of stars, so, apart from perhaps the occasional supernova explosion, detecting real optical variations was considered to be impossible, and therefore little serious attention was paid to Deutsch's report. A few years later, while compiling photoelectric measurements during preparation of the first *Reference Catalog of Bright Galaxies* (de Vaucouleurs & de Vaulcouleurs, 1964), Antoinette de Vaucouleurs independently noticed that fluctuations in the photoelectric magnitudes of NGC 3516, NGC 4051, and NGC 4151 obtained in 1958 significantly exceeded the normal photometric errors (see discussion in de Vaucouleurs & de Vaucouleurs 1968).

The exciting modern era of AGN studies begins with the identification of the first high-luminosity AGNs (Schmidt 1963). Even before Schmidt's discovery that the so-called "quasi-stellar objects" were at high redshift, Sharov & Efremov (1963) had discovered on archival plates taken from 1896-1960 at the Sternberg Astronomical Institute that 3C 273 varied by 0.7m over that period. Special plates they took in the spring of 1962 revealed smaller variations with amplitudes 0.2-0.3m lasting a few days. Matthews & Sandage (1963) discovered from observations they had made going back to 1960 that 3C 48 varied by more than 0.4 magnitudes in the V band over a 13-month period.

Smith & Hoffleit (1963a,b) studied the variability of 3C 273 over the period 1887-1963 using over 600 plates from Harvard plate archives. They found evidence for a 10-year "cyclicity" with a peak-to-peak amplitude of around 0.4 magnitudes and "occasional flashes" of up to a magnitude "lasting about a week."

Only a couple of years later, Dent (1965) discovered variations in the radio flux in flat-spectrum sources.

Less than a decade after the discovery of optical variability of AGNs, X-ray variability was discovered from observations made by the OSO-7, *Uhuru* and *Copernicus* satellites (Davison et al. 1975; Winkler & White 1975)

## 3  VARIABILITY BASICS – SINGLE WAVEBAND VARIABILITY

If we first consider just one waveband, then perhaps the most fundamental questions are: (1) how much does the output vary? and (2) how rapidly does it vary? The answers to both questions tell us fundamental things about the region or regions of the AGN producing the observed emission.

The *amplitude* of variability gives us an idea of the relative importance of variability. It tells us how much the emission from the varying region varies and/or what fraction of the output is contributed by the varying region. A couple of stellar examples will illustrate this. In a supernova explosion the observed luminosity varies enormously. This tells us that the varying region dominates the observed emission and the mechanism responsible for the variability of this region is the main energy production mechanism. On the other hand, the optical variability of our sun on a timescale of minutes to days is very small. Therefore the variations either come from small regions, or the whole sun is only varying by a very small amount. In either case the variability of the sun on short timescales is mostly irrelevant for understanding the fundamental mechanism producing the bulk of the sun's radiation in the optical region. In the radio and X-ray spectral regions the sun is highly variable, however, and this variability *does* tell us important things about the regions producing optical and X-ray emission.

The *timescale* of variability, $\Delta t$, gives information about the rate at which the region varies and it gives an upper limit to the ratio of the size of the region, $L$, to velocity, $v$, at which changes propagate. This timescale is:

$$\Delta t \sim L/v$$

If one knows the speed at which the changes propagate then $\Delta t$ gives the approximate size of the varying region. Alternatively, if one independently knows the size of the varying region then $\Delta t$ tells the speed at which the changes propagate.

One obviously important speed is the speed of light, $c = 3 \times 10^5$ km s$^{-1}$. With only a few exceptions a region cannot vary on a timescale shorter than the light-crossing timescale, $\Delta t \sim L/c$. One scenario in which this limit can be violated is when there is external irradiation of a region. If the re-emitting region is perpendicular to both the irradiation and to the observer then there is no limit on how large the region can be. Another important scenario is when the emitting region is moving relativistically towards the observer. The radiation is then preferentially beamed towards the observer. This is important for BL Lac objects and blazars (Blandford & Rees, 1978).

Two other speeds of particular interest are the orbital speed, $v_{\rm orb} \sim \sqrt{(GM/R)}$, and the sound speed, $v_{\rm s} \sim \sqrt{(kT/m)}$. The orbital speed depends on the distance from the blackhole and the mass of the blackhole, but the fastest observed speeds of the Doppler broadened Fe K$\alpha$ lines are $\sim 3.6 \times 10^4$ km s$^{-1}$ (see Fabian et al. 2000). The sound speed is $\sim 20\ T_4^{1/2}$ km s$^{-1}$ where $T_4 = T/10^4$ K, so

$$c > v_{\rm orb} \gg v_{\rm s}$$

One very important size is the Schwarzschild radius, $R_s = 2GM/c^2$. For the $\sim 10^8$ $M_o$ black hole in a bright AGN this is $\sim 10^{14}$ cm so the relevant timescale is $\sim 1$ hr. The Eddington limit

$$L_{Edd} = 1.3 \times 10^{38} \, (M/M_o) \text{ erg s}^{-1},$$

gives a relationship between $M$ and the luminosity, $L$, if the accretion rate is at the Eddington limit, and hence it also give a relationship between $R_s$ and $L$, and thus between a minimum timescale of variability, $t_{min}$, and $L$ because the timescale of variability must not be less than the light crossing time of the Schwarzschild radius. This minimum variability timescale in seconds is given by

$$\log t_{min} = \log L - 43.1$$

where $L$ is in ergs s$^{-1}$ (Elliot & Shapiro, 1974).

## 4 THE TIME-AVERAGED SPECTRAL ENERGY DISTRIBUTION

The time-averaged spectral energy distribution (SED) over the X-ray to optical region is remarkably similar for AGNs of differing radio types and luminosities. The apparent differences are mostly due to differences in the reddening (see Gaskell et al. 2004). Despite much work over several decades, the origin of the shape of the SED is not understood and this lack of understanding is one of the biggest gaps in our knowledge of how AGNs work.

Broadly speaking, the overall continuum shape from the far-IR to the X-ray region can be characterized very roughly by a power-law:

$$F_\nu \propto \nu^{-1}$$

The SED of AGNs is strikingly different from the SED of a star. A stellar SED is effectively a black body; the SED of an AGN is certainly not.

An explanation of the shape of the SED *must* also be consistent with the variability properties. Despite our poor understanding of the details of how the continuum is produced, the application of simple theoretical considerations to the observed time-averaged continuum gives us important information.

If we take the observed SED to be a sum of black bodies at differing temperatures then we can come up with a relationship between temperature and area (and hence size). The X-ray emission comes from a region several $R_s$ across (light hours for a $10^8$ $M_o$ black hole) while there can be major contributions to the IR emission from a region thousands of $R_s$ across (up to hundreds of light days).

## 5 FUNDAMENTAL QUESTIONS

Probably the most fundamental AGN question is "how is the energy produced?" Since most, but not all, of this energy is seen as electromagnetic radiation, this question becomes, "how are the different continua produced?" A question that then follows is, "are the various continua related?"

We believe that variability probably poses the biggest challenge to understanding how AGNs work. A theory must not only explain the steady-state spectrum of an AGN, but *it must also be able to explain how and why the continua vary*.

Some more specific observational questions concerning variability include:
- What is the amplitude of variability in the various wavebands?
- How are the amplitudes related to the timescales? ("Power Density Spectrum", PDS)
- What are the timescales of variability? What are the *shortest* timescales? What are the *longest* timescales? Are there preferred timescales?
- Is variability periodic?
- Is there evidence for non-linear behavior?
- Are the variations chaotic?
- How is variability of the various continua related?
- How does the variability in various wavebands vary with luminosity?
- Are mean variability properties the same for different classes of AGN? (e.g., radio-loud/radio-quiet, face-on/edge-on, NLS1s/BLS1s)
- Do AGNs of the same class have the same variability properties? (i.e., "do quasars have different personalities?")
- Can the variability properties of an AGN change with time? ("are AGNs moody?")

## 6  OPTICAL AND UV VARIABILITY

### 6.1  The Optical/UV Amplitude is Large

It is important to recognize that optical and ultraviolet variability is enormous! It is enormous in both a relative and an absolute sense. In a typical AGN the annual mean in the V band will typically vary by a few tenths of a magnitude. Even variations of a few hundredths of a magnitude (something we find in essentially every AGN we look at) mean that for a $10^{45}$ ergs/sec AGN the energy equivalent of $\sim 10^{10}$ solar luminosities is switching on and off! The variations are also enormous in a relative sense. In Fairall 9 the UV continuum varied by a factor of over 30 in 180 days (Recond-Gonzalez, 1997). The UV continuum of NGC 4151 varied by a similar factor over a couple of years (Ulrich, Maraschi, & Urry 1997). So clearly, with these enormous factors in the variability, we are dealing with a massive change in the fundamental energy generation mechanism, not some additional superficial phenomenon.

### 6.2  The UV-Optical Continuum Varies as a Unit.

Another important thing to recognize is that the UV and optical continuum varies as a unit. As has been noted above, there is little if any change in the shape of the X-ray to optical spectral energy distribution while the luminosity varies from object-to-object by many orders of magnitudes. This would lead us to suspect that the spectral energy distribution might be the same in a given object as it varies, and this is indeed the case. Although the shape of the optical-UV continuum always appears to "harden" when an AGN brightens, in Fairall 9 the optical spectral shape *remained unchanged* while the intensity increased by a factor of 20 (Lub & de Ruiter 1992).

While to a first approximation one can think of the UV-optical continuum varying as a unit, we will see below that there are now some cases where there are lags across the spectral region, and there is also some evidence that the energy distribution of outbursts varies (Doroshenko et al. 2001).

### 6.3 Timescale of Optical/UV Variability

After the *amplitude* of variability, which tells us whether variability is energetically important or not, probably the most important thing is the *timescale* of variability.

On the longest timescales our knowledge is limited by how far back the observations go. For 3C 273 we have photographs going back to the 1880s (Smith & Hoffleit 1963). We have observations of NGC 4151 from 1906 to the present (Lyutyi & Oknyanskij 1987). In both cases there is considerable variability, and we would have been unaware of the greatest outbursts without the historical record. In both cases also there is variability on the timescales of decades. In NGC 4151 there is variation on the longest timescale with a 80-year "period." There are also quasi-periodic variations on timescales of years to a decade or so. We will discuss the important question of possible periodicities below. The long timescale variations are of large amplitude. Because of starlight contamination, the real amplitude will certainly be larger than it appears to be in the optical.

Variations on shorter timescales have smaller amplitudes (see discussion of the power-density spectrum below), and are therefore harder to detect. This is an area where modern high-precision photometry has a lot to contribute. *X-ray* variability on timescales as short as less than an hour is commonly observed in some objects. Similarly rapid (intra-night) optical variability (sometimes called "microvariability") is well established in optically-violently variable (OVV) objects. Jang & Miller (1997) found intra-night variability to be more common in radio-loud AGNs than radio-quiet ones. There have been conflicting reports of microvariability in Seyfert galaxies. Because of the difficulty of measuring microvariability, observational errors might be contributing to some differences in reports, but Merkulova (2000, 2002) concludes that intra-night variability is really transient in character and manifests itself with different probabilities for different galaxies.

### 6.4 Power Density Spectra

The power density spectrum (PDS, P(*f*)) potentially contains information about the nature of variability. For example, a "shot-noise model" where variations arise from a stochastic series of independent overlapping events will produce a so-called "red noise" power spectrum of the form

$$P(f) \propto f^{-\alpha}$$

with α ~ 2 that becomes "white noise" with α = 0 for low frequencies. Red noise is characteristic of many astrophysical and terrestrial systems. Many processes in nature produce so-called "1/*f* noise" (see Bak 1996)

Kunkel (1967) using 100 day bins found that for 3C 273 α ~ 2 from 0.12 to 1.83 cycles/yr. Collier & Peterson (2001) find that on timescales τ~5-60 days, the mean UV and optical PDSs for 13 AGNs are equivalent. The combined UV/optical PDS has α = 2.13. For sources with measured X-ray PDS indices, they find that the optical/UV and X-ray PDSs are indistinguishable. They present evidence that higher mass systems have larger characteristic timescales.

### 6.5  Variations are Logarithmic

Optical astronomers almost invariably plot the brightness of AGNs in magnitudes. In a magnitude plot, the light curves of AGNs look symmetric both under time-reversal and if the magnitude scale is inverted. The distribution of magnitudes is roughly normal, so the distribution of fluxes must therefore be roughly lognormal. Lyutyi & Oknyanskij (1987) made the important discovery that there was a linear relationship between the variations in the U-band flux ($\Delta F_U$) and the U-band flux ($F_U$) itself for NGC 4151. This suggested that the amplitude of optical variability was directly proportional to the optical flux of the AGN. We will show below that both lognormality of variations and a linear relationship between flux and variability also hold for X-ray variations.

### 6.6  No True Periodic Variations

A wide range of phenomena in astronomy are periodic (orbits, pulsations, etc.) and these periods provide important physical information about the systems they arise in. As soon as AGN variability was discovered, an obvious question to ask was, "are the variations periodic?" Smith and Hoffleit (1963) reported a 10-year "cyclicity" in 3C 273, but Kunkel (1967) reported that there were "no outstanding periodicities." For NGC 4151, Lyutyi & Oknyanskij (1987) discuss quasi-periods of tens of days, ~ 4, ~ 14, and ~ 80 years, but found no true periods for more than a few cycles. Longo et al (1996) similarly found no evidence for periodicities.

Mention must be made of the BL Lac object OJ 287. Its long-term light curve, assembled from data accumulated over a century, shows nine nearly evenly-spaced outbursts (Kidger, M. R. 2000). Due to the uneven sampling, an exact value for the period cannot be determined, although the average seems to be ~ 11 years. However, there is predictive power to this finding, as the next maximum should occur in 2006, if the periodicity is real.

## 7  X-RAY VARIATIONS

### 7.1  X-rays Vary A Lot

It has long been appreciated that the amplitude of X-ray variability is large. Terrell (1986) found variations of at least an order of magnitude in the Vela 5B light curve of Cen A going back to 1969. The most spectacular cases of variability are seen in so-called narrow-line Seyfert 1s (NLS1s - see below). For example, during a 30-day monitoring of IRAS 13224-3809, Boller et al. (1997) found five giant-amplitude "flares", the largest with an amplitude of 60. In this object there is no evidence for a non-variable component (Gaskell 2004). As with UV and optical variations, we can say that the variations represent changes in the fundamental energy mechanism.

### 7.2 X-rays Vary Rapidly

Variations in the X-ray region are more rapid than optical/UV variations of similar amplitude. In IRAS 13224-3809, for example, there was a factor of 2 variation in about 20 minutes (Boller et al. 1997). PHL 1092 has shown a flux increase of almost a factor of four in less than an hour. If these sorts of observed changes are interpreted as changes in isotropic flux, they imply radiative efficiencies that exceed the maximum that can be achieved from a rotating black hole. This suggests that the emission is not isotropic and there is boosting due to relativistic motions (see Boller & Brandt 1999 for details). Certainly the regions varying must be within 15 $R_s$ or (less likely) be smaller regions further out. The important thing to note is that variations are taking place on a *light-crossing timescale* and not a viscous timescale. This probably rules out the standard accretion disk model of Shakura & Sunyaev (1973, 1976).

### 7.3 Soft X-rays Vary The Most

The most impressive X-ray variations, such as those mentioned in the previous section, occur in the soft X-rays. The soft component was found to vary the most in several studies of Seyfert 1s (Nandra et al. 1997; Turner et al. 1999; Markowitz & Edelson 2001). This could be due to the presence of a softer continuum emission component varying more strongly than a harder component or to a nonconstant single component that becomes softer as the source becomes brighter. In NLS1s the fractional variability is independent of the waveband (Edelson et al. 2002). This could be because the hard component is relatively weak in NLS1s.

### 7.4 X-ray Power Density Spectra

Considerable effort has been put into determining X-ray power density spectra. EXOSAT data showed that X-ray variability is scale-invariant "red noise" from timescales of minutes to days (Lawrence et al. 1987; McHardy & Czerny 1987). Lawrence & Papadakis (1993) found that the PDS had a mean power law index, α ~ 1.55 and pointed out that this mean slope is inconsistent with both standard shot-noise processes and traditional '1/f noise'. This sort of PDS was noted to be similar to PDSs of galactic black hole X-ray binaries, although on a much longer timescale. This naturally raises the possibility that there could be similarities in the processes causing variability and this led to the search for other similarities in the PDSs.

Edelson & Nandra (1999) obtained a high-quality PDS for NGC 3516, which showed a progressive flattening of the power-law slope from 1.74 at short timescales to 0.73 at longer timescales. This gave a characteristic variability timescale corresponding to a cutoff temporal frequency of about a month. This is about six orders of magnitude longer than is seen in stellar mass galactic black hole sources and thus suggested that the timescale scales with the mass. Similar breaks in the PDS have been found in a number of other objects, but Uttley, McHardy & Papadakis (2002) found no low-frequency break in NGC 5548.

### 7.5 X-ray Variations Are Logarithmic

Just as Lyutyi & Oknyanskij (1987) discovered that optical variability is proportional to the mean optical flux level, Uttley & McHardy (2001) similarly discovered that the X-ray variability of the stellar mass black hole Cyg X-1 and the accreting milli-second pulsar SAX J1808.4-3658 was linearly related to the flux level. They also suggested that AGNs could show a similar relationship but they were only able to compare pairs of states of slightly differing mean luminosity in three AGNs. Gaskell (2004) showed that there is indeed a linear relationship between X-ray variability and X-ray flux (see Fig. 1 below) and argued that this flux-dependent behavior of the variability rules out linear shot-noise models. Vaughan, Fabian &. Nandra (2003) found a similar relationship for MCG-6-30-15.

As noted above optical light curves are approximately lognormal. Gaskell (2004) shows that for IRAS 13224-3809 the large variations in both the soft X-ray flux observed by ROSAT and the hard X-ray flux observed later by ASCA can be well-fit by a two-parameter lognormal distribution (see Fig. 2).

Although at first glance the variations of the *ASCA* light curve for IRAS 13224-3809 appear to exhibit non-stationary behavior with quiescent low-states and more active flaring high states, our results show that the multiplicative variance is constant. Monte Carlo simulations of constant $\sigma_{mult}$ give excellent matches to the observed X-ray light curve without the need to invoke special low and high states. This supports a picture in which the long-term variability is fundamental.

A lognormal distribution of X-ray fluxes suggests that the emitting regions could have a lognormal size distribution or the energies could have a lognormal distribution.

### 8 RELATIONSHIP BETWEEN X-RAY AND OPTICAL CONTINUA

Because of the large amplitude and rapidity of X-ray variability it has been commonly considered to be the driving variability of AGNs. But is this really so? There are two models that venture to explain the AGN processes producing lags. In one model, the UV-optical bands observed are the result of reprocessed X-rays. The primary X-rays heat the cooler matter, perhaps lying in the disk or torus, which then re-radiates the reprocessed radiation. In this case, the prediction is that the optical follows the X-rays. In another model, the X-rays are Comptonized UV-optical photons. In this scenario, a corona of relativistic electrons Comptonizes the UV-optical radiation, thereby producing X-rays. This model predicts that the X-rays follow the UV-optical emission.

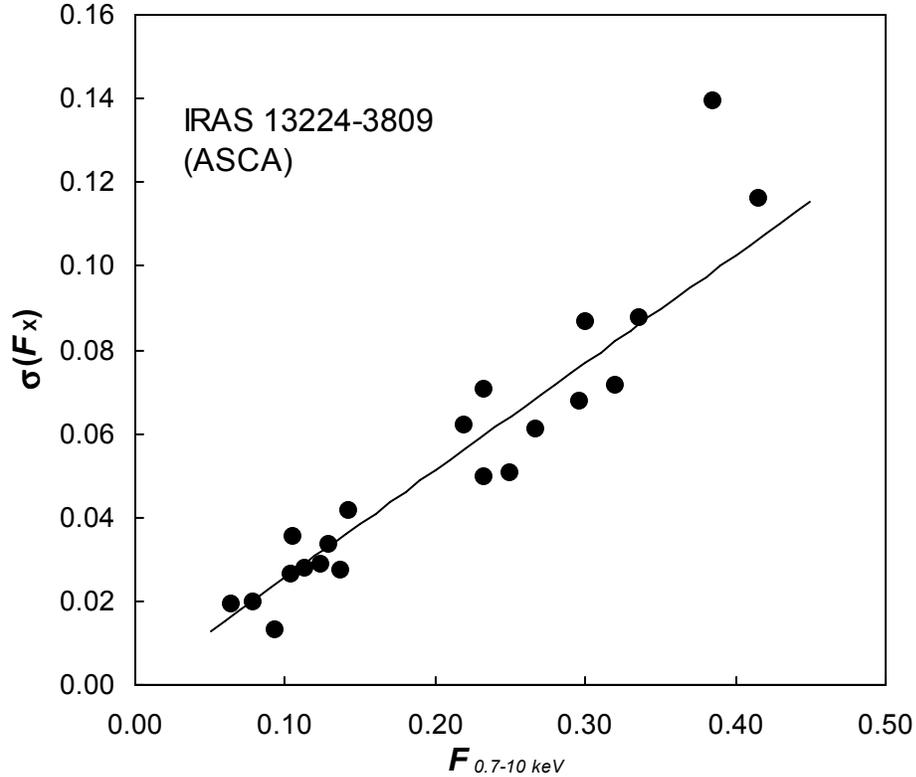

FIGURE 1  The standard deviations versus the mean count rate in half-day bins for IRAS 13224-3809 (from Gaskell 2004). Both axes are in ASCA counts over the 0.7 – 10 keV energy range. The fit line is $\sigma(F_X) = 0.256\, F_X$.

While each model has its own specific prediction, the observational results have been unclear or insufficient to rule out any theory. The first lag comparison was done by Lyutyi (1978) who found that X-ray variability, while correlated with optical variability both on long and short timescales, has a greater amplitude and shorter timescale. He found that the optical band may have lagged that of the X-rays. Since then many researchers have used simultaneous observations in different wavebands to search for any kind of lag between two spectrum regimes. A range of results have been found for various objects. For example, Done et al. (1990) showed that there was very little optical variability (<1%) on timescales of days in NGC 4051 and no apparent correlation with the much larger amplitude X-ray variability. On a longer timescale Peterson et al. (2000) suggested that the optical and X-ray light curves for NGC 4051 were correlated on timescales of months−years. The conclusion from the *International AGN Watch* campaign focused on NGC 4151 (Edelson et al. 1996) was that there was no clear relationship between any of the wavebands. In studying NGC 3516 there was "no significant correlation or simple relationship" (Edelson et al. 2000).

However, in another study of NGC 7469, the X-rays actually seemed to follow the UV by about five days (Nandra et al. 2000). A six-year study of the RXTE light curve of NGC 5548 (Uttley et al. 2003) shows that the X-ray light curve is strongly correlated with the optical light curve on long (~1 yr) timescales.

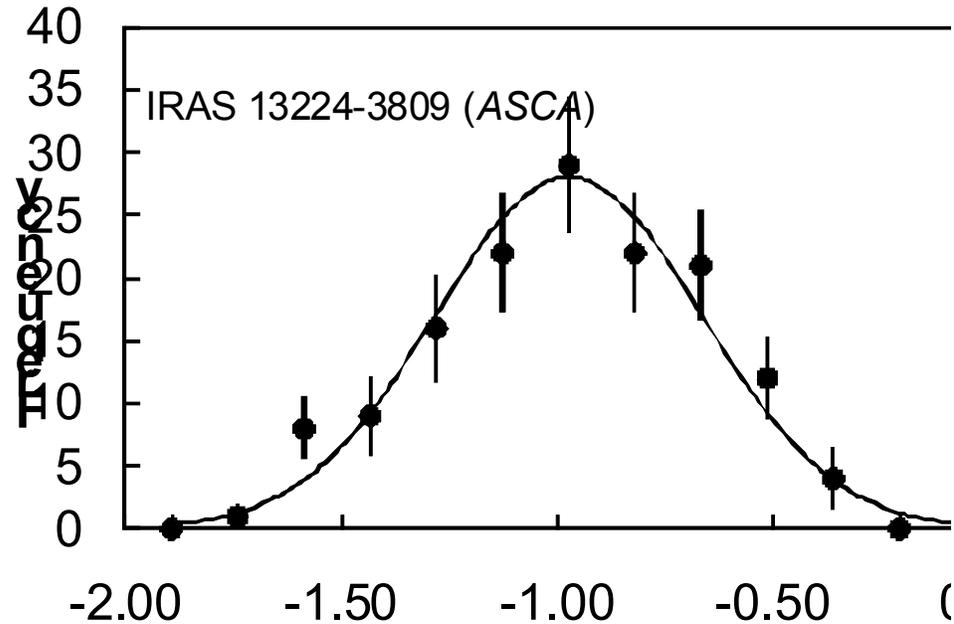

FIGURE 2   The frequency distribution of the logarithms of the ASCA count rates over 5 ksec intervals (from Gaskell 2004). The curve is a Gaussian of standard deviation 0.435 dex.

In a more recent international observing campaign targeting Ark 564 (see Fig. 3), the optical band followed the X-rays by about 1.5 days (Shemmer et al. 2001; see also Doroshenko et al. 2006). For 3C 390.3 we find that the optical lags the X-rays by 4.5 days (see Fig. 4 below, and further discussion in Gaskell 2006) Clearly the area of X-ray correlations is one in which much more work is needed.

Uttley et al. (2003) find that the amplitude of the long-term optical variability in NGC 5548, after accounting for the host galaxy contribution, is *larger* than that of the X-ray variability. In our own study (Gaskell et al. in preparation) of the radio-loud AGN 3C 390.3 (see Fig. 5 above), we find that our optical fluxes show the same amplitude of variability *even without allowing for host galaxy contamination.* This means that the fractional amplitude of the daily optical variations is *larger* than the fractional amplitude for the X-ray observations. These results are of great importance because they rule out X-ray reprocessing as the main source of the optical/X-ray correlation.

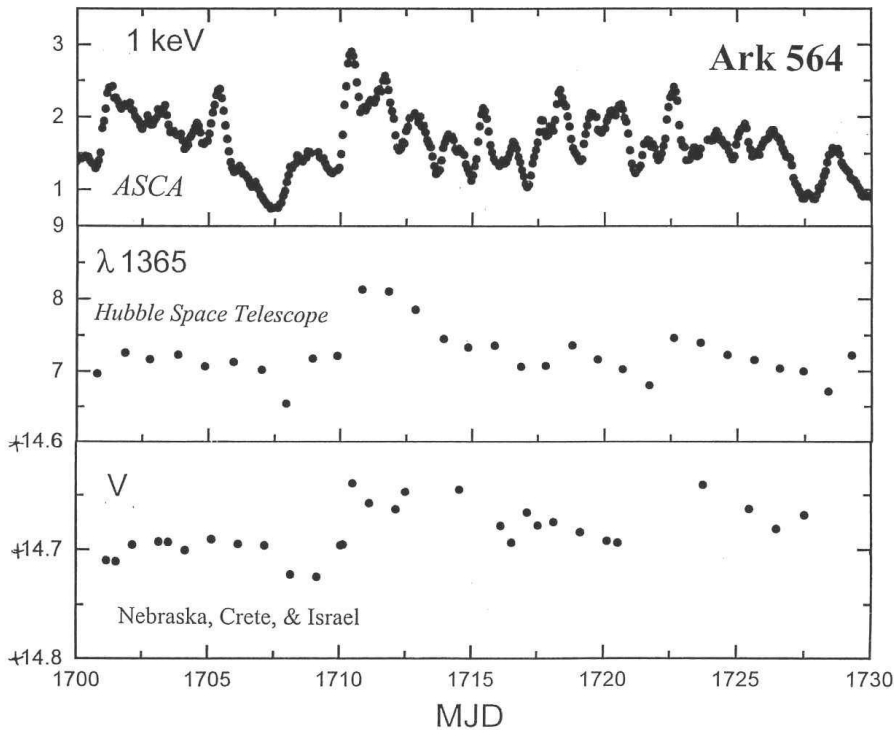

FIGURE 3   Multi waveband light curves of Ark 564. Data from Shemmer et al. (2001) and Doroshenko et al. (2006).

## 9 WAVELENGTH DEPENDENT LAGS IN UV/OPTICAL

Lags across the UV/optical waveband are predicted by most models and have long been sought. Such a lag was convincingly found for the first time by Collier et al. (1998) in NGC 7469 (see also Kriss et al. 2000). The continuum at 7000Å lagged the continuum at 1400Å by about 1.5 days. Collier (2001) found a 1.4 day (rest-frame) lag across the optical passband for the gravitationally-lensed quasar 0957+561. Recently, Oknyanskij et al. (2002) have found a similar delay across the optical passband in NGC 4151. Their result is very interesting because our earlier International AGN Watch campaign had put an upper limit on the UV/optical lag of < 0.15 days (Edelson et al. 1996). This means that *the geometry of the X-ray/optical emission regions has changed over a period of several years.* Clearly more studies of more objects and repeated studies of objects are needed, and some such studies have been instigated as a joint campaign between a number of observatories in the Former Soviet Union and the University of Nebraska.

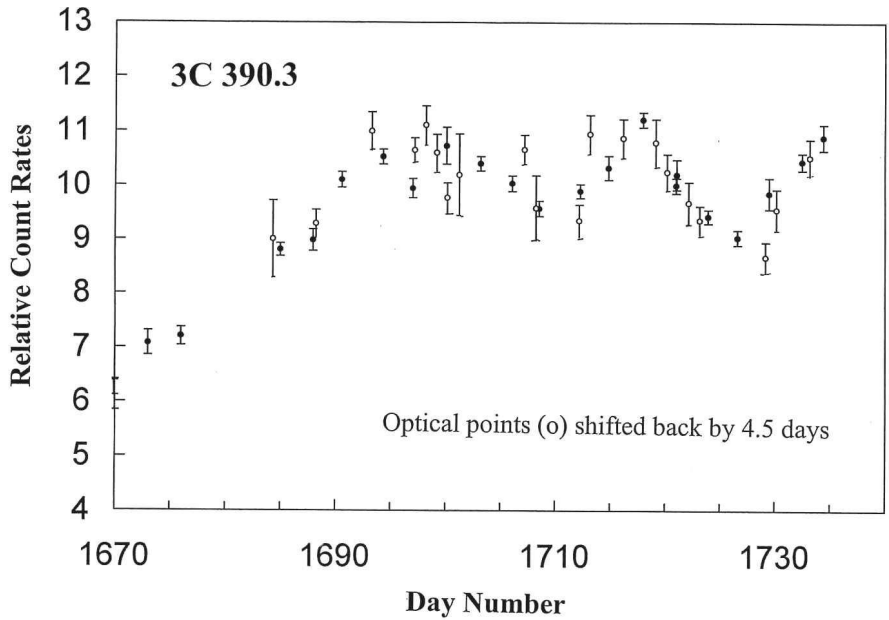

FIGURE 4    Simultaneous RXTE (filled circles) and unpublished University of Nebraska optical (open circles) light curves for 3C 390.3. Note that no galaxy component has been subtracted from the V-band measurements and that the optical points have been shifted back by 4.5 days to emphasize the agreement with the X-ray light curve.

It is important to note that the lags across the UV/optical region and the upper limits on such lags are on the *light-crossing time* timescale. This means that whatever is causing the connection between different wavebands is propagating at close to the speed of light. This is commonly suggested to be external illumination, but as noted elsewhere in this review, there are very serious problems with having all optical radiation arising from reprocessing. We therefore suggest that the linking factor might not be irradiation but accelerated particles striking matter, as happens in the solar chromosphere. Differences in local geometry, viewing angle, and direction of motion relative to the line-of-sight could then explain the diversity of X-ray/optical relationships described in section 8.

## 10    NON-LINEARITY AND NON-STATIONARITY

One obvious way to try to explain the structure of light curves is by a linear superposition of discrete events. Such modeling has been attempted by Fahlman & Ulrych (1975, 1976), Scargle (1981), and others. However, there is much evidence that X-ray and optical light curves are both non-linear and non-stationary. Angione & Smith (1985) pointed out that the 3C 273 light curves of Smith (1965) and Terrell & Olsen (1970) "scarcely seem to refer to the same object." Vio et al, (1991) argued that the optical variations of 3C 345 are non-linear and also non-stationary. A multi-fractal analysis performed by Longo et al. (1996) clearly indicated non-linear intermittent behaviour in the long term (1910 – 1991) B-band light curve of NGC 4151. Leighly & O'Brien (1997) showed that the flares and quiescent periods in the light curve of 3C 390.3 suggest that its X-ray variability is nonlinear. The non-stationary character of the light curve could, however, be evidence that the variability power spectrum has not turned over at low frequencies. They suggested that the character of the variability is similar to that seen in Cygnus X-1, which has been explained by a reservoir or self-organized criticality model. Green, McHardy, & Done, (1999) showed that NGC 4051 is also not statistically stationary on timescales of ~ 1 year. The non-stationarity and non-linearity of light curves means that the power spectra do not adequately represent all the information contained in the light curve. Models in which the variability events are correlated rather than random are needed to describe the observed light curves. Gaskell (2004) points out that the lognormal nature of variability (see above) needs to be considered when evaluating stationarity. There is no evidence for IRAS 13224-3809 that the *multiplicative* variance is not stationary.

## 11 CHAOS?

A power-law PDS can theoretically arise from a chaotic system where global coherent variability is described by three or more non-linear differential equations showing deterministic chaos. Such. However Czerny, & Lehto (1997) have analyzed the EXOSAT light curves of eight AGNs and find no signs of deterministic chaos. In half the AGNs the variability is clearly of a stochastic nature, and in the other half the variability was not strong enough to determine its character, but stochastic variability was again favored.

## 12 SELF-ORGANIZED CRITICALITY?

Vio et al. (1991) suggested that AGNs could be self-organized critical (SOC) systems, i.e., they become organized into a state where they are on the edge of instability. A pile of sand is a classic example of such an SOC system – the addition of a few grains of sand can cause a major avalanche. One of the signatures of an SOC system is that it produces power-law distributions (Bak 1996). Negoro et al. (1995) argued that observational features in the light curve of Cygnus X-1 strongly suggest the presence of numerous reservoirs with different capacities for triggering X-ray fluctuations, a key assumption of the model based on the self-organized criticality. Xiong et al. (2000) have produced SOC models for accretion disk fluctuations. Their model can produce light-curves and power-spectra for the variability that agree with the range observed in optical and X-ray studies of AGN and X-ray binaries.

Despite the promise of SOC models we note that the lognormal flux distribution and the constancy of $\sigma_{mult}$ (Gaskell 2004) are incompatible with the power-law distribution of flaring amplitudes expected from simple SOC behavior. Takeuchi et al. (1995) offer a more detailed model that could fit the X-ray behavior better.

## 13 NARROW-LINE SEYFERT 1 GALAXIES (NLS1s)

Narrow-line Seyfert 1 galaxies (Gaskell 1984, Osterbrock & Pogge 1985) are so called because the central engine with its surrounding dense gas can be seen directly, as in Seyfert 1 galaxies, but the permitted optical emission lines arising from gas (the broad-line region) are much narrower than in normal Seyfert 1 galaxies. NLS1s show the most extreme X-ray variability. Boller, Brandt, & Fink (1996) showed that, as a class, NLS1s also have strong soft X-ray excesses and greater X-ray variability than would be expected for their luminosity. At a particular X-ray luminosity, the excess variance is typically an order of magnitude larger for NLS1s than for Seyfert 1s with broad optical lines (Leighly 1999a). IRAS 13224-3809 mentioned above is a NLS1.

The enhanced excess variance exhibited by NLS1s can be interpreted as evidence that they are scaled-down versions of broad-line objects, having black hole masses roughly an order of magnitude smaller and requiring an accretion rate an order of magnitude higher (Leighly, 1999a). It is possible, though, that the X-ray variability of NLS1s has higher amplitude flares than normal broad-line Seyfert 1s (non-NLS1s) and might be different.

Boller, Brandt, & Fink (1996) showed that NLS1s show a strong soft X-ray excess and a steeper X-ray spectrum. The strength of the soft excess is correlated with the variability parameters, so that objects with strong soft excesses show higher amplitude variability (Leighly 1999b). It is important to understand why this is so.

Since NLS1s show extreme X-ray variability it is reasonable to ask whether they also show extreme optical variability. Young et al. (1999) looked unsuccessfully for optical microvariability in the extremely X-ray variable AGN IRAS 13224-3809, but Miller et al. (1999) did find significant microvariability on one night for the same object.

We have carried out a long-term multi-observatory international study of the variability of Ark 564 (Shemmer et al. 2001; Doroshenko et al. 2006). We have found a number of rapid events on intra-night timescales that we have observed at more than one observatory. Some of the events correlate with X-ray events (see Fig. 4), but the optical fractional amplitudes are much less than the X-ray fractional amplitudes. It is hard to say whether such rapid low-amplitude events are also common in non-NLS1s because of the lack of a suitable control sample. Combining our observations with earlier ones from Doroshenko we now have coverage of Ark 564 for over a decade a half (Doroshenko et al. 2006) and it shows long-term variations similar to non-NLS1s.

We have also completed a large-scale optical photometric study of additional NLS1s searching for variability from intra-night timescales to timescales of years (Klimek et al. 2004). Despite looking on ~ 40 nights, we have not detected significant intra-night variability. We do not see the sort of 0.3 mag intra-night variability Miller et al. (1999) reported for IRAS 13224-3809. The lack of suitable control samples again makes it hard to say how the level of microvariability we find compares with that of non-NLS1s, but we can confidently say that:

(i) NLS1s do *not* show the sort of extreme variability in the optical that they show in the X-ray region, and

(ii) The amplitudes of intra-night optical variability for NLS1s are not *significantly* greater than for non-NLS1s.

On longer timescales the optical variability of NLS1s seems to be similar to that of non-NLS1s, but again the lack of a control group of non-NLS1s is a problem. We mention the lack of control group problem because studies of non-NLS1s have been biased towards objects that are known to vary. The effect of this bias needs to be evaluated. We need to know whether AGNs of the same class (e.g., radio-quiet non-NLS1s) have different personalities.

## 14  BEAMING

In this review we are intentionally avoiding known beamed sources such as BL Lacs and other OVV AGNs. There are definite differences between the variability properties of OVVs and "normal" AGNs (see review by Ulrich, Marschi, & Urry 1997), especially in amplitude and timescale, but we do wonder whether these differences have been over-emphasized, particularly since OVVs and non-OVVs tend to be observed by different observers and discussed at different meetings. If OVV light curves are appropriately scaled it is not clear how different they are from non-OVV light curves. Since explaining the observed optical/X-ray amplitudes and correlations in non-OVVs necessarily requires the transmission of large amounts of energy in relativistic particles, we suggest that beaming could well be a major factor in "normal" AGNs. For example, the non-linear intermittent behavior found by Longo et al. (1996) in NGC 4151 led them to suggest that the physical mechanism responsible for the variability of NGC 4151 could be similar to the mechanism responsible for the variability of the OVV 3C 345.

## 15  DO RADIO-LOUD AND RADIO-QUIET AGNs VARY THE SAME WAY?

If there is beaming going on in non-OVVs then the AGNs most likely to be similar to OVVs are radio-loud AGNs. This raises the question, "do radio-loud and radio-quiet AGNs vary the same way?". We have pointed out above that after allowing for reddening and dust properties in AGNs (Gaskell et al. 2004) the continua of radio-loud and radio-quiet AGNs appear to be very similar in the UV and the optical.

OVV AGNs are all radio-loud, and because their extreme variability is believed to be due to relativistic beaming (see previous section), this has led to a belief that radio-loud AGNs are more variable than radio-quiet ones. We believe, however, that apart from OVVs there is little compelling evidence for this. The historical amplitudes of variability of the radio-loud AGN 3C 273 (which is sometimes classified as a blazar) are similar to those of well-studied radio-quiet Seyfert galaxies such as NGC 4151. We have given above examples of where the continuum of radio-quiet Seyfert galaxies has varied by over a factor of 20.

We have found that a recently discovered radio-quiet AGN, PDS 456 (Torres et al. 1997), surprisingly displays as much optical variability as a comparable radio-loud object. PDS 456 is the most luminous object in the "local" universe and is similar to the luminous quasars seen when the universe was only 10-20% of its current age. Strongly X-ray variable (Reeves et al. 2000), this object is comparable in luminosity to 3C 273, the classic radio-loud bright nearby AGN. We found that the total range of optical variation in PDS 456 was 30% over a span of about 120 days. In comparison, 3C 273, a comparable radio-loud object, has a typical seasonal range roughly half of this at 16%. In fact, 75% of the seasons during which 3C 273 was observed have a variation range of less than PDS 456's 30%. On only one occasion in 30 years did 3C 273 vary as much as PDS 456.

This similarity of the variability of PDS 456 to 3C 273 suggests to us that the variability mechanisms of radio-loud and radio-quiet AGNs are the same. We intend to make further observational studies to investigate this.

## 16  VARIABILITY-LUMINOSITY DEPENDENCE

Paltani & Courvoisier (1997) looked at the luminosity dependence of variability in IUE spectra. The UV variability amplitude goes as L to the –0.08 and they found that an index of –0.5 is definitely excluded. This luminosity dependence has no natural explanation in terms of discrete events.

In the X-ray region Barr & Mushotsky (1986) showed that the timescale of variability is correlated with the luminosity and Wandel & Mushotzky (1986) showed that there was a corresponding relationship between mass and the timescale of variability. The amplitude of X-ray variability of large samples of radio-quiet AGNs shows that the variability amplitude scales inversely with luminosity (Green, McHardy & Lehto 1993; Lawrence & Papadakis 1993; Nandra et al. 1997). For NLS1s, time series analysis shows that the excess variance from the NLS1 light curves is inversely correlated with their X-ray luminosity (Leighly 1999). However, with a logarithmic slope of ~-0.3, the dependence of the excess variance on luminosity is flat compared with broad-line objects and the expected value of -1 from simple models.

## 17  AN OVERALL PICTURE

A lot of work has been done researching AGN variability, and the field is far from slowing down. Many fundamental questions remain unanswered, and new ones keep arising the more these objects are studied, but so far there is not enough evidence to pick out a winning AGN model, if indeed one even as of yet exists. Conflicting findings, perhaps arising from poor sampling, or unusual "moods" of objects, are observational reasons for this. On the theoretical side, no theory has so far been able to account for all the properties AGNs are observed to have.

We believe that a theory must explain the following:

- The rapidity of AGN variability.

- That variability of AGNs is fundamental in that it is related to the main energy-generation process.

- That soft X-ray variability dominates energetically.

- The relationship of the optical band to the X-rays. Studies of this relationship tell us that the optical emission is not simple reprocessing, although reprocessing to some degree is probably going on. The complexity and variety of temporal relationships between the X-rays and the optical even in the same object needs to be explained.

- That the optical variability of radio-loud and radio-quiet AGNs is quite similar.

- That it is hard to distinguish between beamed and non-beamed sources on the basis of many variability characteristics.  This suggests that the mechanisms producing beamed and non-beamed variability are similar and perhaps the same.

Although we are far from having a complete theory, we believe a new picture is emerging:

A. We believe that the rapidity and amplitude of variability make the "standard" model of a quasar powered by viscous dissipation in a relatively stable accretion disk (Shakura & Sunyaev, 1973) untenable.

B. We believe instead that electromagnetic processes must dominate. We suggest that the dominant process underlying all variability is relativistic flares.

## 18   CONCLUDING REMARKS:  THE IMPORTANCE OF INTERNATIONAL COLLABORATION

In AGN research, the challenges faced by equipment limitations, weather, and simply getting enough telescope time have contributed to the lack of quality optical (and IR) data that can solidly support or reject various theories. Uninterrupted continuous coverage of a range of AGN in all wavebands is desired but unrealistic.  The closest we can get to this ideal situation is through international collaborations and coordination with X-ray and UV observations during the operational lifetimes of orbiting astronomical satellites.  It cannot be stressed enough how important these multinational observing campaigns have been and are for contributing to the needs of the astronomical and scientific world.

There are a number of practical issues that make collaborative efforts essential.  Good sampling is needed in order to cover just about every type of timescale, including minutes, hours, days, weeks, months, years, and even decades.  For microvariability, which is on the smallest timescales of hours or less, the variations are of such low amplitude that systematic instrumental errors start to become a problem.  In this case independent confirmation by other observers strengthens the validity of the observations, thereby helping to remove the doubt surrounding the reality of any microvariability claim.  In order to get full 24-hour coverage of an object, we need observatories that are spread out in longitude.  Additionally, having observers in different geographic locations is helpful in overcoming weather problems and also helps overcome the problem that observers seldom get enough telescope time at any one observatory.  The longest timescales require us to look at the different historical archives observatories have built up.

Satellites can be taken for granted.  They seem to incessantly stream down a wealth of data such that we cannot obtain from the ground.  While satellites might seem to tirelessly pump out invaluable data, they do not live forever. We need to be making the best use of all resources while they are available to us. For example, the Vela 5B satellite provided a good X-ray light curve of Cen A continuously on a daily basis from 1969-1979. Today, the RXTE satellite, for example, will only be in operation for a few more years. The finite lifetime of X-ray missions means that *now* is an important time to coordinate optical monitoring with the X-ray observations. RXTE is currently observing a number of AGNs for the long term, providing us the opportunity to build up a decent sized archive of simultaneous optical-X-ray data that can be used in correlation studies. There are other satellites to take advantage of, such as the Chandra, XMM-Newton, and INTEGRAL observatories.

In order to get the best optical coverage, observations need to be taken around the world through collaborations between as many researchers as possible. One of the sad things about the study of variability over the last couple of decades has been the inferiority of optical coverage compared to X-ray coverage. Almost no X-ray light curve has a comparable simultaneous optical light curve. Yet, optical data comes much cheaper than the X-ray data! One X-ray point might cost $10,000 or even much more, while optical data can ring up at as little as $10 per point. We hope this situation will be rectified in the future.

*Acknowledgements*

MG wishes to thank the American Astronomical Society and US National Science Foundation through grant AST 9980200 for making possible his participation in the 2002 Euro-Asian Astronomical Society Meeting in Moscow, and he wishes to express his thanks to the Sternberg Astronomical Institute of Moscow State University, the Special Astrophysical Observatory, and the Crimean Astrophysical Observatory for their generous hospitality. We would also both like to thank the University of Nebraska Research Council for financial support of this research. Additional support for our AGN research at the University of Nebraska has been received from NASA, the US National Science Foundation through grant AST 03-07912, the Space Telescope Science Institute, the Howard Hughes Foundation, Nebraska EPSCoR, University of Nebraska Area of Strength funding, and UCARE/Pepsi-Cola. We are grateful to Valya Doroshenko for useful comments on this paper.